\pdfoutput=1

\documentclass[conference]{IEEEtran}

\usepackage{balance}  
\usepackage{graphics} 
\usepackage{graphicx}
\usepackage{times}    
\usepackage{url}      
\usepackage{algorithm}
\usepackage[noend]{algpseudocode}
\usepackage{caption}
\usepackage{subcaption}

\makeatletter
\def\url@leostyle{%
  \@ifundefined{selectfont}{\def\UrlFont{\sf}}{\def\UrlFont{\small\bf\ttfamily}}}
\makeatother
\def\pprw{8.5in}
\def\pprh{11in}

\usepackage[pdftex]{hyperref}

\newcommand{\argmax}{\textit{argmax}}

\begin{document}

\title{Crowdsensing in Opportunistic Mobile Social Networks: A Context-aware and Human-centric Approach}

\author{\IEEEauthorblockN{Phuong Nguyen}
\IEEEauthorblockA{Department of Computer Science\\
University of Illinois at Urbana-Champaign\\
}
{\small \tt{pvnguye2@illinois.edu}}
\and
\IEEEauthorblockN{Klara Nahrstedt}
\IEEEauthorblockA{Department of Computer Science\\
University of Illinois at Urbana-Champaign\\
}
{\small \tt{klara@illinois.edu}}
}

\maketitle

\begin{abstract}
In recent years, there have been efforts to collect human contact traces during social events (e.g., conferences) using Bluetooth devices (e.g., mobile phones, iMotes).  The results of these studies have enabled the ability to do the crowd-sourcing task from within the crowd, in order to answer questions, such as: what is the current density of the crowd, or how many people are attending the event?  However, in those studies, the sensing devices are usually distributed and configured in a certain manner.  For example, the number of devices is fixed, people register for the devices on a volunteering basis.  In this paper, we treat the above problem as an optimization problem and draw the connection to the vertex cover problem in graph theory.  Since finding the optimal solution for minimum vertex cover problem is NP-complete, approximation algorithms have to be used.  However, we will show that the well-known approximation algorithms do not perform well with the crowd-sensing task.  In this paper, we propose the notions of node observability and coverage utility score and design a new context-aware approximation algorithm to find vertex cover that is tailored for crowd-sensing task.  In addition, we design human-centric bootstrapping strategies to make initial assignment of sensing devices based on meta information about the participants (e.g., interests, friendship).  The motivation is to assign the sensing task to a more ``socialized'' device to obtain better sensing coverage.   We perform comprehensive experiments on real-world data traces obtained from previous experimental studies in conference and academic social context.  The results show that our proposed approach significantly outperforms the baseline approximation algorithms in terms of sensing coverage.
\end{abstract}

\section{Introduction}\label{sec:intro}
Crowd-sourcing has changed the way people obtain needed services, new ideas, or content by soliciting contributions from a large group of people, and especially from an online community (e.g., Amazon Mechanical Turks).  There are also efforts of using crowd-sourcing to monitor, obtain information about the crowd -- which is referred to as crowd-sensing \cite{chaintreau2007impact, hui2005pocket, pietilanen2012dissemination}.  For example, during Olympic London 2012, organizers released a smartphone app that allowed users to upload their location information to help determine how to manage the crowds and the associated city resources.  With the proliferation of mobile devices and communication technologies, it has become more possible to use technologies to understand the behavior of large crowds.  For example, during social events, such as conferences, some attendants are given Bluetooth-enabled devices (e.g., mobile phones, iMotes) to collect opportunistic contacts (i.e., the Bluetooth scanning results of neighboring devices).  The collected contacts, or traces, can be used to answer a variety of questions about the crowd.  For example, what is the density of the crowd, how many people, or groups of people exist in the crowd (i.e., for crowd monitoring purpose), or what is the current opportunistic contact graph between devices (i.e., to understand the ability of data dissemination within the crowd).  We refer to the tasks of answering these questions as the crowd-sensing tasks.  

There have been previous experiments on collecting opportunistic mobile traces, such as SIGCOMM'09 \cite{pietilanen2012dissemination, crawdad} or UIM \cite{vu2010joint}, where the collected sensing data are valuable for studying crowd-sensing tasks.  However, in those experiments, the settings are usually configured in a certain manner.  For example, the number of sensing devices (e.g., Bluetooth-enabled devices that periodically scan for neighboring devices) is fixed, devices are given to attendants on a volunteer basis, and the sensing interval (i.e., the interval during which opportunistic contacts are collected) is fixed.  In addition, in those experiments, each sensing device does the sensing task in isolation and data from all sensing devices are only collected at the end of the experiment.  As a result, such data could not help answer timely questions about the crowd.  Besides, while crowd sensing task can be handled successfully, given all individual devices collecting sensing data all of the time, such approach is overly demanding of the device's resources, and not energy-efficient.  In this paper, we propose a crowd-sensing model, in which the sensing nodes periodically connect to a centralized server to send the collected data and receive the instruction for the next sensing interval.  In addition, we treat the number of sensing devices and the length of sensing interval as the given constraints, and try to optimize the assignment of sensing task to appropriate devices to maximize the sensing coverage.  


Under this scheme, we will show that the crowd-sensing task poses some similarities to sensor placement task \cite{meguerdichian2001exposure}, where the problem is to find an optimal set of positions to place sensors in order to obtain the best coverage of the environment, given the constraint as the number of available sensors.  However, crowd-sensing problem is more challenging due to the spatio-temporal and social nature of the interactions between people in the crowd.  For example, the collected sensing data might be highly overlapped if people carrying sensing devices are nearby each other at the same location.  In another example, since social relationships and personal interests might have influence on who people frequently meet or interact with, the centralized server should take such ``out-of-band'' information into its assignment of sensing task to devices.

In this paper, we show the connection between the crowd-sensing problem with the vertex cover problem in graph theory.  While finding the optimal solution in vertex cover is NP-complete, we show that the constrained versions of existing approximation algorithms, such as random-based approximation, or greedy approximation \cite{hochbaum1982approximation, dinur2005hardness}, can be used to derive approximate solutions.  However, since those algorithms are designed for generic graph, they do not take into account the spatio-temporal and human-centric characteristics of the crowd-sensing task and optimize the individual coverage of covering vertex instead of combined coverage, as in greedy approximation.  To solve this problem, we propose a new context-aware approximation algorithm that is tailored for crowd-sensing task.  Particularly, we propose the notion of \textit{node observability} and \textit{coverage utility score} to optimize the combined sensing coverage objective.  In addition, we incorporate the out-of-band, human-centric information about the participants, such as personal interests, and social relationship, to improve the bootstrapping of the crowd-sensing task.  The experimental results on real-world mobile traces show a significant improvement in sensing coverage while satisfying the optimization constraints.

In summary, our contributions in this paper are as follow:
\begin{itemize}
\item We model the crowd-sensing problem as an optimization problem and draw the connection to the vertex cover problem in graph theory.
\item We propose a general 2-stage framework for incorporating vertex cover approximation into crowd-sensing task.
\item We propose the notions of node observability and coverage utility score and design a new context-aware approximation algorithm to find vertex cover that is tailored for crowd-sensing task.
\item We perform comprehensive experiments on real-world data traces to verify the effectiveness of our proposed approach.
\end{itemize} 

The paper is organized as follow: in Section~\ref{sec:probdef}, we describe the sensing model and formally define the problem.  In Section~\ref{sec:vertexcover} we show how to solve the crowd-sensing problem using minimum vertex cover approximation.  In Section~\ref{sec:contextualapprox}, we describe our proposal of a context-aware approximation algorithm for crowd-sensing task.  We show the results of experiments on real-world datasets in Section~\ref{sec:experiments} and discuss the related work in Section~\ref{sec:relatedwork}.  Finally, in Section~\ref{sec:conclusions}, we conclude the paper and discuss some future work.

\section{Crowd-sensing model and problem definition}\label{sec:probdef}

In this section, we first describe the crowd-sensing model before formally defining the problem.

\subsection{Crowd-sensing model}

\begin{figure}[!h]
\centering
\includegraphics[width=0.8\columnwidth]{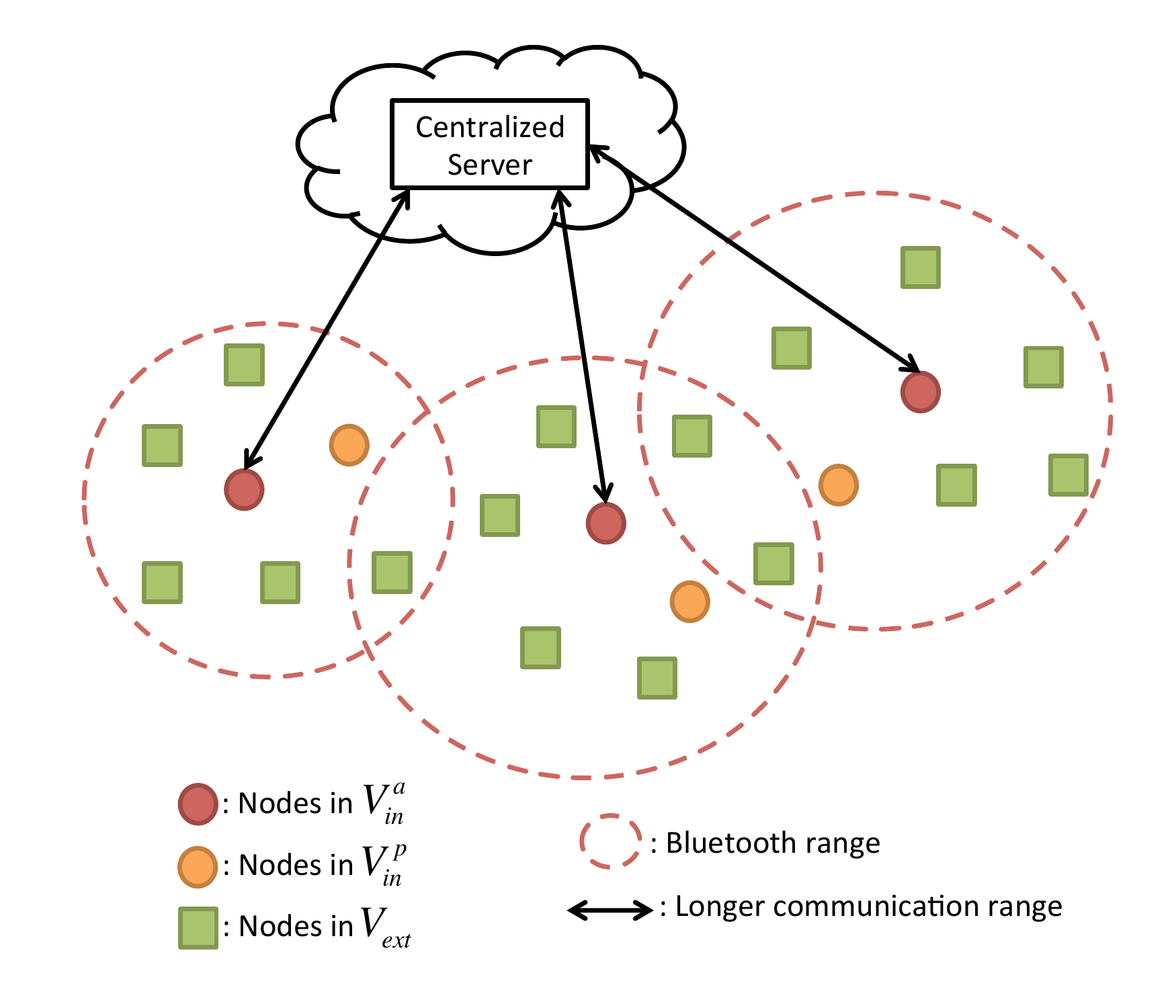}
\caption{Crowd-sensing model with centralized server and distributed sensing devices}
\label{fig:sensingmodel}
\end{figure}

Let us denote \textbf{V} as the set of devices (e.g., mobile phones) attending a crowd event (e.g., conference sessions, classes on campus, etc.).  Among devices in \textbf{V}, there is a sub-set of devices $\textbf{V}_{in}$ (represented as circles in Figure~\ref{fig:sensingmodel}) that register as the participants of the crowd-sensing experiment -- we call it as the set of \textit{internal devices}.  The remaining set of devices $\textbf{V}_{ex}$ (represented as the green rectangles in Figure~\ref{fig:sensingmodel}) does not participate in the experiment -- we call it as the set of external devices (apparently, we have $\textbf{V} = \textbf{V}_{in} \cup \textbf{V}_{ex}$).  An sensing application is installed on each internal device that runs in background and periodically connects to a centralized server to i) send collected sensing data, and ii) receive sensing instructions.  

In this paper, we refer the sensing task to the task of collecting opportunistic Bluetooth contacts, or wireless contacts, in mobile ad hoc networks.  As a result, we assume that every device in \textbf{V} has its Bluetooth in discoverable mode.  In addition, for the sensing devices, they need to periodically do the scanning of neighboring Bluetooth-enabled devices and record the observed contacts.  Besides, since the sensing devices also need to contact with the centralized server for sending the collected data or receiving instructions, they need to have connection to the server by longer range communications, such as Wifi, or 3G.

Although, ideally, we would like to have all devices in $\textbf{V}_{in}$ perform the sensing task, we argue that such an approach is not energy efficient and even not necessary.  On one hand, performing the sensing task by scanning for neighboring Bluetooth devices is very energy-consuming.  On the other hand, the sensing results of multiple sensing devices could be overlapped and thus, not all the data are helpful in covering the crowd.  The overlapping situation is more serious when the people carrying the sensing devices are at the same place, or move together.  As a result, we only require a subset of registered devices doing the sensing task at a time, and we consider the number of sensing devices as a given constraint.  Particularly, at any point of time, an internal device can be either in \textit{sensing mode} (during which, the device collects the wireless contacts of its neighboring devices), or \textit{non-sensing mode} (during which, the device does not do the sensing task and waits for the sensing instruction from centralized server for the next sensing interval).  We denote the devices that are in the sensing mode as $\textbf{V}_{in}^{a}$ (the red circles in Figure~\ref{fig:sensingmodel}), and the ones that are in the non-sensing mode as $\textbf{V}_{in}^{p}$ (the orange circles in Figure~\ref{fig:sensingmodel}).  Apparently, we have $\textbf{V}_{in} = \textbf{V}_{in}^{a} \cup \textbf{V}_{in}^{p}$.

\begin{figure}[!h]
\centering
\includegraphics[width=0.8\columnwidth]{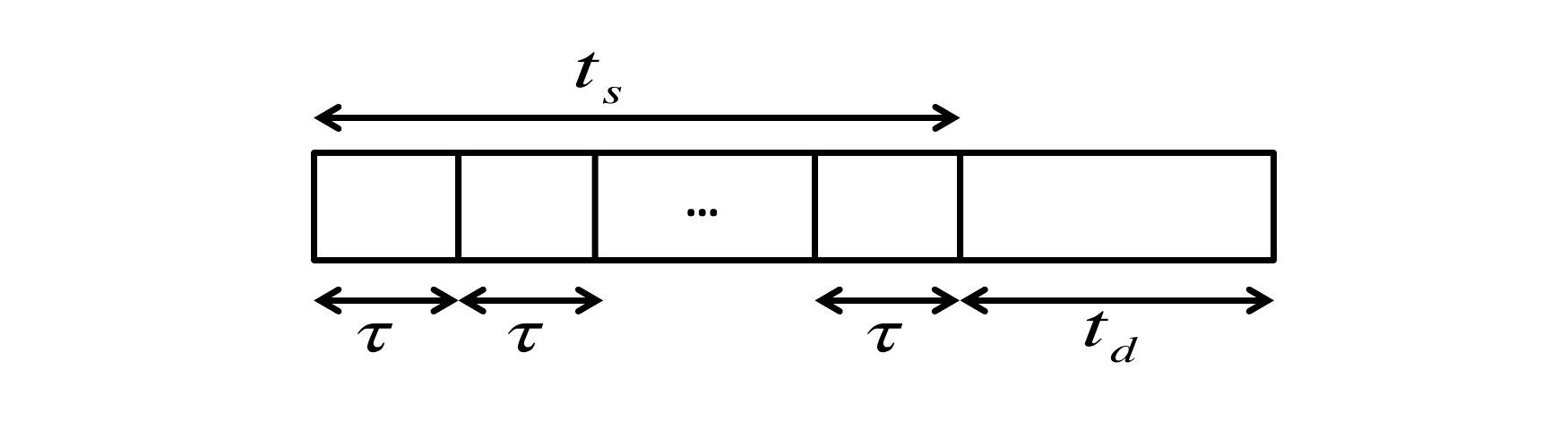}
\caption{Operating time division of an internal device}
\label{fig:timedivision}
\end{figure}

Each internal device operates under two different time windows (Figure~\ref{fig:timedivision}).  During \textit{sensing time window} $t_{s}$, if an internal device is in the sensing mode, it periodically senses the neighboring environment for wireless contacts and stores the data locally before sending to the server at the end of sensing interval.  The value of $t_{s}$ is chosen as a multiplication\footnote{The reasons for not choosing $t_{s} = \tau$ are to reduce the number of times communicating with centralized server, and to increase the chances of observing neighboring contacts, which might be missing if just a single scan is used, due to weak signal or obstacles} of inquiry interval of wireless sensor on each device $\tau$ (i.e., the time gap between two consecutive scanning of neighboring devices): $t_{s} = T \times \tau$, with $T$ is a fixed integer.  If an internal device is in non-sensing mode during $t_{s}$, it simply does nothing.  During \textit{decision time window} $t_{d}$, both sensing and non-sensing devices listen to the instructions from the centralized server to decide which ones would do the sensing task in the next round.  Also during $t_{d}$, the sensing devices will send its locally stored data of contacts observed during previous $t_s$ to the server.  After $t_{d}$, all the devices go to a new $t_{s}$ period with a new set of sensing devices decided by the server. 

\subsection{Problem definition}

Let us denote $\textbf{E}_{t_s}$ as the set of wireless contacts observed by sensing devices during the time interval $\textit{t}_{s}$.  For each sensing device $\textit{v} \in \textbf{V}_{in}^{a}$, let us denote $\textbf{E}_{t_s}(\textit{v})$ as the set of wireless contacts that \textit{v} observes during $t_{s}$.  The set of wireless contacts obtained from devices in $\textbf{V}_{in}^{a}$ can be used to construct a contact graph $\textbf{G}_{t_s} = (\textbf{V}_{t_s}, \textbf{E}_{t_s})$ of devices during $t_s$, with $\textbf{V}_{t_s}$ being the set of nodes seen during $t_s$ (including the sensing nodes) and each edge in $\textbf{E}_{t_s}$ represents a contact between two devices in $\textbf{V}_{t_s}$ during $t_s$.  From now on, we might use nodes/devices, and edges/contacts interchangeably as they refer to the same notion.  

Since $\textbf{V}_{in}^{a}$ is only a subset of $\textbf{V}_{in}$ ($\textbf{V}_{in}^{a} \subseteq \textbf{V}_{in}$), it is possible that the set of observed contacts during $t_s$ might not be complete: $\textbf{E}_{t_s} \subset \overline{\textbf{E}_{t_s}}$, where $\overline{\textbf{E}_{t_s}}$ is the set of contacts observed during $t_s$ if $\textbf{V}_{in}^{a} \equiv \textbf{V}_{in}$.  Therefore, our objective is to find a set of vertices $\textbf{V}_{in}^{a}$, whose size is limited to a predefined $n$, that maximizes the number of observed contacts during $t_s$.  The crowd-sensing problem is thus formally defined as follow: \\

\textbf{\textit{Problem Definition}}: Given a set of internal devices $\textbf{V}_{in} = \textbf{V}_{in}^{a} \cup \textbf{V}_{in}^{p}$, for each sensing time interval $t_s$, find an optimal set of sensing devices $\textbf{V}_{in}^{a*}$, whose size equals a predefined $n$ ($n \leq |\textbf{V}_{in}|$), that maximizes the number of wireless contacts $|\textbf{E}_{t_s}|$ observed during $t_s$: 
$$\textbf{V}_{in}^{a*} = \argmax_{\textbf{V}_{in}^{a} \subset \textbf{V}_{in}, |\textbf{V}_{in}^{a}| = n} \, |\textbf{E}_{t_s}|$$

In the above definition, the absolute number of observed contacts (i.e., $|\textbf{E}_{t_s}|$) represents the coverage ability and is used as the maximization objective.  Equivalently, we can also use the ratio between the number of observed contacts by devices in $\textbf{V}_{in}^{a}$ and that by all devices in $\textbf{V}_{in}$, i.e., $|\textbf{E}_{t_s}| / |\overline{\textbf{E}_{t_s}}|$, to measure the  coverage capability.  We refer this ratio as \textit{sensing coverage ratio}.

\section{Crowd-sensing by vertex cover optimization}\label{sec:vertexcover}

In this section, we show the connection between the crowd-sensing problem and the vertex cover problem \cite{vertexcover}.  We first revisit the vertex cover problem and then, show why finding the minimum cover set of a contact graph also gives us an efficient solution for crowd-sensing problem.  After that, we describe the constrained versions of two approximation algorithms \cite{hochbaum1982approximation, dinur2005hardness} for vertex cover problem and how they can be applied to crowd-sensing problem.

\subsection{Vertex cover of a graph}

\begin{figure}[!h]
\centering
\includegraphics[width=1.0\columnwidth]{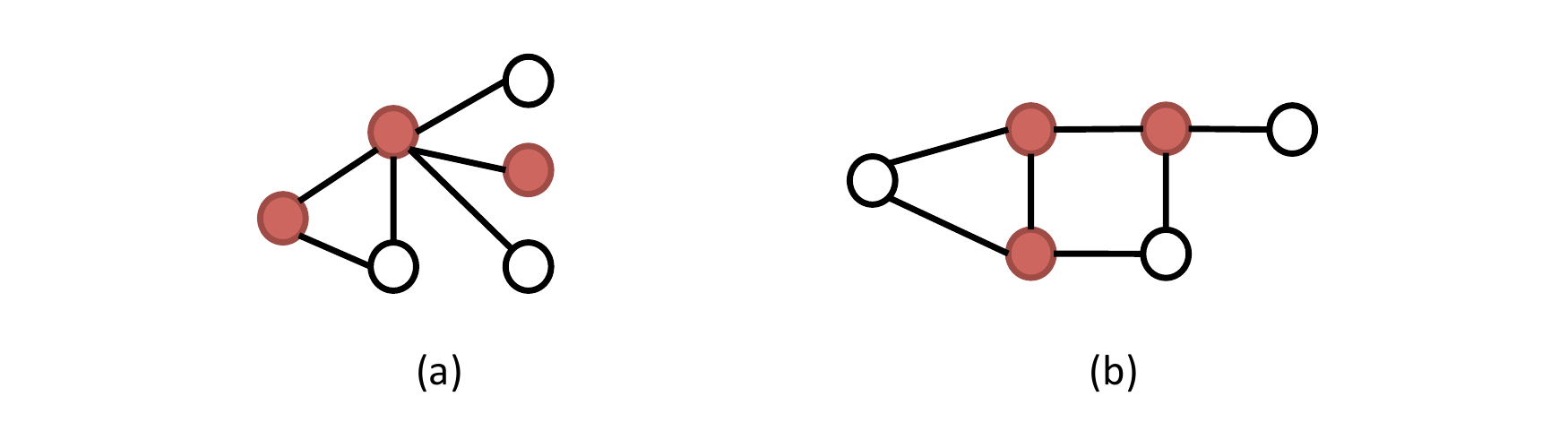}
\caption{Examples of vertex cover in graphs (red nodes are vertex cover)}
\label{fig:vertex-cover}
\end{figure}

A vertex cover of a graph is a set of vertices such that each edge of the graph is incident to at least one vertex of the set.  Formally, a vertex cover of an undirected graph $\textbf{G} = (\textbf{V}, \textbf{E})$ is a subset $\textbf{V}' \subset \textbf{V}$ such that if edge $(u, v)$ is an edge of \textbf{G}, then $u \in \textbf{V}'$, $v \in \textbf{V}'$, or both. It is not difficult to see there might have more than one vertex cover of a graph. In Figure~\ref{fig:vertex-cover}a, the red nodes represent one example of the graph's vertex cover.

A minimum vertex cover is a vertex cover of smallest possible size. In Figure~\ref{fig:vertex-cover}b, the red dots represent an example of the graph's minimum vertex cover.

From here, we can see more clearly the connection between vertex cover problem and crowd-sensing problem.  As every edge in a graph is incident to at least one vertex in vertex cover, the set of nodes in the vertex cover could essentially ``observe'' all contacts (i.e., edges) of the graph.  In crowd-sensing scenario, the vertex cover is basically the set of sensing nodes that can observe all number of contacts between all nodes in the graph.  In addition, since the number of sensing devices in crowd-sensing problem is constrained to a predefined number $n$ (which is preferred to be small, due to the cost and high energy consumption of having too many sensing devices), it is also desirable to find the minimum vertex cover as the set of sensing devices.  In case the size of minimum vertex cover is greater than $n$, we need to exclude vertices from the cover set in a way that minimizes the effect to the observable edges (i.e., to make fewest number of edges become non-observable by vertices in the vertex cover).  One simple strategy is to exclude the vertices that have smallest number of edges connected to it. 

Finding minimum vertex cover of a graph, however, is a NP-complete problem \cite{karp1972reducibility}.  There have been efforts to come up with approximate solutions \cite{hochbaum1982approximation, dinur2005hardness}.  In the next section, we discuss the constrained version of the two well-known approximation algorithms for minimum vertex cover problem with our modification to comply with the constraint on the maximum number of sensing nodes $n$.

\subsection{Approximation algorithms for vertex cover problem}

The first approximation algorithm is based on random selection of vertices into vertex cover set.  This randomization strategy is similar to the 2-Approximation algorithm that finds a factor-2 approximation by repeatedly taking both endpoints of an randomly selected edge into the vertex cover \footnote{The 2-Approximation algorithm is not directly applicable for crowd-sensing task, because not always the two vertices of a chosen edge are the internal devices}.  In our random-based approximation algorithm (Algorithm~\ref{alg:topkrandomapprox}), we needs to comply with the constraint on the number of vertices on the cover set as well as the fact that the selected vertices for the sensing task needs to be in $\textbf{V}_{in}$.  Particularly, we repeatedly randomly select a vertex from the set of internal devices and add it to the cover set (while also removing all adjacent edges to the selected vertex), until the size of the cover set equals $n$.

\begin{algorithm}
\caption{Top-n Random-based Approximation Vertex Cover}\label{alg:topkrandomapprox}
\begin{algorithmic}[1]
\Procedure{TopNRandomApproxCover}{$\textbf{V}_{t_s}$, $\textbf{E}_{t_s}$, $\textbf{V}_{in}$, $n$}
\State $\textbf{V}_{in}^{a} \gets \emptyset$
\While {$\textbf{E}_{t_s} \neq \emptyset$ and $|\textbf{V}_{in}^{a}| \leq n$}
	\State pick any $u \in \textbf{V}_{in}$
	\State $\textbf{V}_{in}^{a} = \textbf{V}_{in}^{a} \cup \{u\}$
	\State delete all edges incident to either $u$ from $\textbf{E}_{t_s}$.
\EndWhile \textbf{end while}
\State return $\textbf{V}_{in}^{a}$
\EndProcedure
\end{algorithmic}
\end{algorithm}

In the second approximation algorithm, i.e., Top-n Greedy Cover (Algorithm~\ref{alg:topkgreedy}), we repeatedly select a vertex in $\textbf{V}_{in}$ with highest node degree from the contact graph to add to the vertex cover.  The motivation behinds this approximation algorithm is that the higher the node degree of a vertex is, the more incident edges it has, and thus the more likely the selected vertices in the vertex cover can fully cover all edges in the contact graph.  Similar to Algorithm~\ref{alg:topkrandomapprox}, this greedy algorithm also ends when the size of cover set reach the limiting number $n$, to comply with the constraint on the number of vertices on the cover set.

\begin{algorithm}
\caption{Top-n Greedy Vertex Cover}\label{alg:topkgreedy}
\begin{algorithmic}[1]
\Procedure{TopNGreedyCover}{$\textbf{V}_{t_s}$, $\textbf{E}_{t_s}$, $\textbf{V}_{in}$, $n$}
\State $\textbf{V}_{in}^{a} \gets \emptyset$
\While {$\textbf{E}_{t_s} \neq \emptyset$ and $|\textbf{V}_{in}^{a}| \leq n$}
	\State select $u \in \textbf{V}_{in}$ with highest degree
	\State $\textbf{V}_{in}^{a} = \textbf{V}_{in}^{a} \cup \{v\}$
	\State delete all edges incident to $u$ from $\textbf{E}_{t_s})$
\EndWhile \textbf{end while}
\State return $\textbf{V}_{in}^{a}$
\EndProcedure
\end{algorithmic}
\end{algorithm}

In the following section, we will describe a framework to plug-in the above approximation algorithms to solve crowd-sensing problem.

\subsection{Crowd-sensing by vertex cover approximation}

At the beginning of crowd-sensing task, since the centralized server does not even have a contact graph to start with approximation algorithms, the assignment of sensing task to a set of vertices $\textbf{V}_{in}^{a} \in \textbf{V}_{in}$ is done by a \textit{bootstrapping} algorithm.  Then, after each sensing interval $t_s$, using the sensing data collected by nodes in $\textbf{V}_{in}^{a}$, the centralized server can construct a contact graph between devices of the previous interval and use this as the input of the approximation algorithms (e.g., Top-n Random-based Approximation Vertex Cover or Top-n Greedy Vertex Cover -- Algorithm~\ref{alg:topkrandomapprox} and \ref{alg:topkgreedy}) to find the new vertex cover for the next iteration.

The general framework of applying vertex cover approximation to crowd-sensing problem is described in the following 2-stage algorithm (Algorithm~\ref{alg:crowdsensing-vc}):

\begin{algorithm}
\caption{Crowd-sensing by Vertex Cover Approximation}\label{alg:crowdsensing-vc}
\begin{algorithmic}[1]
\Procedure{CrowdSensingVertexCover}{$\textbf{V}_{in}$, $\textbf{V}_{ex}$, $n$, $rounds$}
\State $\textbf{V}_{in}^{a} \gets \emptyset$
\State $currentRound \gets 0$
\While{$currentRound < rounds$}
	\If {$currentRound = 0$}
		\State $\textbf{V}_{in}^{a} \gets bootstrap(\textbf{V}_{in})$
	\Else
		\State $\textbf{V}_{in}^{a} \gets vertexCoverApprox(\textbf{V}_{t_s}, \textbf{E}_{t_s}, \textbf{V}_{in}, n)$
	\EndIf
	\State $(\textbf{V}_{t_s}, \textbf{E}_{t_s}) = getCoveredGraph(\textbf{V}_{in}^{a})$
	\State $currentRound \gets currentRound + 1$
\EndWhile \textbf{end while}
\EndProcedure
\end{algorithmic}
\end{algorithm}

In Algorithm~\ref{alg:crowdsensing-vc}, $bootstrap(\textbf{V}_{in})$ is the bootstrapping method that returns the initial assignment of sensing task to a subset of vertices $\textbf{V}_{in}^{a}$ (sized $n$) in $\textbf{V}_{in}$. $vertexCoverApprox(\textbf{V}_{t_s}, \textbf{E}_{t_s}, \textbf{V}_{in}, n)$ is the call to a vertex cover approximation algorithm (e.g., Top-n Greedy, or Top-n Random-based Approximation Vertex Cover algorithms) to find the vertex cover sized $n$ given a contact graph (i.e., $(\textbf{V}_{t_s}, \textbf{E}_{t_s})$) and the set of registered devices (i.e., $\textbf{V}_{in}$). The method $getCoveredGraph(\textbf{V}_{in}^{a})$ constructs a contact graph given a vertex cover set based on all the wireless contacts that vertices in the cover set observe during a sensing interval. $rounds$ represents the number of rounds to run the crowd-sensing task.

\section{Context-aware approximation and human-centric bootstrapping for crowd-sensing}\label{sec:contextualapprox}

While both of the above approximation algorithms are simple and seemingly intuitive, they avoid important underlining information about the contact graph and how it is constructed.  Particularly, since each node in the graph is actually a device carried by a \textit{human user}, the opportunistic contacts between devices are influenced by the spatio-temporal interactions between human users carrying the devices and the social context of the crowd event.  For example, the collected opportunistic contacts might be highly overlapped (and thus, is not very helpful in covering a large crowd) if the sensing devices move together, or stay at the same location.  In another example, in the context of an academic conference, people tend to attend presentation sessions of the topics of their interests.  As a result, a sensing node tends to observe surrounding devices of people sharing similar interests.  These insights motivate us to design a new \textit{context-aware} approximation algorithm and \textit{human-centric} bootstrapping methods for the crowd-sensing task that take into account the spatio-temporal interactions between devices and the social context of the crowd.

In this section, we describe in more details our proposals of approximation algorithm and bootstrapping strategies as parts of the two stages in the general crowd-sensing framework (Algorithm~\ref{alg:crowdsensing-vc}).  We first describe our proposed human-centric bootstrapping strategies based on meta information of the participants (i.e., $\sim bootstrap(\textbf{V}_{in})$ -- the first stage of the framework).  After that, we introduce the notions of node observability and coverage utility score that are the key components of the proposed context-aware approximation algorithm (i.e., $\sim vertexCoverApprox(\textbf{V}_{t_s}, \textbf{E}_{t_s}, \textbf{V}_{in}, n)$ - the second stage of the framework). 

\subsection{Human-centric bootstrapping strategies}

As mentioned in the previous section, at the beginning of the crowd-sensing process, without a contact graph to start with, we need to bootstrap the process by assigning the sensing task to a subset of nodes sized $n$ in $\textbf{V}_{in}$.  

While the most basic way of bootstrapping is to randomly select nodes as sensing nodes, we propose to use ``out-of-band'' meta information about the social relationships and interests of human users who carry the internal devices to initialize the selection of sensing nodes.  Particularly, we consider two types of relationship between people: \textit{friendship} (i.e., whether two or more people know each other in person) and \textit{interests} (i.e., whether two or more people share the same personal interest).  Our approach is motivated from the fact that human mobility and interaction are influenced by their interests and how they are connected socially.  Specifically, people who are friends or share similar interests tend to get together more often than the ones who do not know each other, or do not have anything in common.  As a result, if we know such relationships between people, we can appropriately assign the sensing task to the persons who have a lot of friends, or the ones who share common interest with a lot of people. 

The three bootstrapping strategies used in this paper are summarized as follow:

\begin{itemize}
\item \textit{Random-based bootstrapping}: Randomly select a subset of $n$ devices from $\textbf{V}_{in}$ as the sensing devices.
\item \textit{Friendship-based bootstrapping}: Build a friendship social network of participants in $\textbf{V}_{in}$.  Sort all participating people in descending order of their node degree in friendship social network.  Select top-$n$ people as the initial set of users carrying sensing devices.
\item \textit{Interest-based bootstrapping}: Group people into groups of interests. Sort all the interest groups in descending order of its number of members.  For each group in the top-$n$ groups, select a member with the most diverse affiliation (i.e., the member with the most number of groups he/she belongs to).
\end{itemize}

In this paper, the fact that the proposed bootstrapping strategies are limited to only using friendship and personal interests is due to the availability of such meta information in the experimenting datasets.  However, we believe that these are the good examples of using out-of-band information in bootstrapping the selection of sensing nodes.

\subsection{Node observability and coverage utility score}

Before describe our proposed context-aware approximation algorithm for finding vertex cover, we introduce the notions of node observability and coverage utility score that are the key components of the proposed algorithm.

An appropriate utility function is essential for any approximation algorithm.  For example, in Top-n Greedy Vertex Cover algorithm, the utility function is actually the method that calculates the node degree of each internal vertex in the contact graph.  The motivation of the greedy objective function is that the higher node degree is, the better coverage a node has.  While such a motivation seems to be intuitive, it only tries to obtain the local optimal and does not capture the fact that the coverage by different nodes might be overlapped and some nodes are less visible than others.  As a result, the combined coverage by multiple sensing nodes might not be good enough to cover all contacts in the crowd.  

To account for different levels of visibility between nodes, we propose the notions of \textit{node observability} for non-sensing nodes that helps measure how ``easy'' a non-sensing node can be observed by other sensing nodes.  In addition, to establish an objective function that can help optimize the overall sensing coverage (i.e., the global optimal), we propose the notion of \textit{coverage utility score} for sensing nodes.  This utility score accounts for how important the contacts that a sensing node observes are, in terms of contributing to a high overall sensing coverage. 

Specifically, node observability is defined as the number of sensing nodes that observe a given non-sensing node during a sensing interval.  The higher the observability score is, the more visible a node is during that interval. \\

\textit{Definition 1 (Node observability)}: During a sensing interval $t_s$, node observability $\sigma_{t_s}(v)$ of a non-sensing node $v \not\in \textbf{V}_{in}^{a}$ is defined as: 
$$\sigma_{t_s}(v) = |\{u \,|\, u \in \textbf{V}_{in}^{a}, (u, v) \in \textbf{E}_{t_s}\}|$$

Since the ultimate objective is to obtain the best combined coverage -- to cover as many contacts between nodes as possible (even the contacts with less visible nodes), covering nodes with low observability is also very important.  In addition, if a node already has a high observability, the fact that this node is observed by some other sensing nodes is not very important (as that means overlapped contacts, or the node is in an area with a redundant number of sensing nodes). Based on this, we propose a notion of coverage utility score that gives higher reward to a node that can observe less visible nodes in its sensing results. \\

\textit{Definition 2 (Coverage utility score)}: Coverage utility score $\Delta_{t_s}(u)$ of a sensing node $u \in \textbf{V}_{in}^{a}$ during a sensing time interval $t_s$ is defined as: 
$$\Delta_{t_s}(u) = \Sigma_{v | (u, v) \in \textbf{E}_{t_s}, v \not\in \textbf{V}_{in}^{a}} \sigma^{-1}_{t_s}(v)$$


\subsection{Context-aware approximation algorithm}

In the second stage of the crowd-sensing algorithm (Algorithm~\ref{alg:crowdsensing-vc}), after receiving data from the sensing nodes, the centralized server update the set of sensing nodes by using an approximation algorithm (i.e., $vertexCoverApprox(\textbf{V}_{t_s}, \textbf{E}_{t_s}, \textbf{V}_{in}, n)$).  We now describe our proposed context-aware approximation algorithm (Algorithm~\ref{alg:contextawareapprox}) based on the notions of node observability and coverage utility score introduced in the previous section.

In our approximation algorithm, we first calculate the observability scores of all non-sensing nodes and coverage utility scores of the sensing nodes from the collected sensing data.  These scores are used to assess the effectiveness of the sensing nodes (i.e., using coverage utility scores) and the sensing potential of the non-sensing nodes (i.e., using observability scores).  Particularly, top $k$ ($k < n$) nodes with highest coverage utility scores are kept in the set for the next sensing interval.  The $(n-k)$ nodes with lowest coverage utility scores are replaced by non-sensing nodes with highest observability scores.

\begin{algorithm}
\caption{Context-aware Approximation Algorithm}\label{alg:contextawareapprox}
\begin{algorithmic}[1]
\Procedure{ContextAwareApprox}{$\textbf{V}_{t_s}, \textbf{E}_{t_s}, \textbf{V}_{in}, n$}
\State $\textbf{V}_{in}^{a'} \gets \textbf{V}_{in}^{a}$ \Comment Copy the current set of sensing nodes
\State $\textbf{V}_{in}^{a} \gets \emptyset$ \Comment Empty the new set of sensing nodes
\For {each $v \in \textbf{V}_{in} \setminus \textbf{V}_{in}^{a'}$} \Comment Calculate node observability
	\State $\sigma_{t_s}(v) \gets |\{u \,|\, (u, v) \in \textbf{E}_{t_s}\}|$
\EndFor
\For {each $u \in \textbf{V}_{in}^{a'}$} \Comment Calculate coverage utilities
	\State $\Delta_{t_s}(u) \gets \Sigma_{v | v \not\in \textbf{V}_{in}^{a'}, (u, v) \in \textbf{E}_{t_s}} \sigma^{-1}_{t_s}(v)$
\EndFor
\State Sort $\textbf{V}_{in}^{a'}$ by $\Delta_{t_s}$ in ascending order
\State Add top-$k$ nodes in $\textbf{V}_{in}^{a'}$ to $\textbf{V}_{in}^{a}$
\State Sort $\{\textbf{V}_{in} \setminus \textbf{V}_{in}^{a'}\}$ by $\sigma_{t_s}$ in descending order
\State Add top-$(n-k)$ nodes in $\{\textbf{V}_{in} \setminus \textbf{V}_{in}^{a'}\}$ to $\textbf{V}_{in}^{a}$
\State return $\textbf{V}_{in}^{a}$
\EndProcedure
\end{algorithmic}
\end{algorithm}  

In the above algorithm, $k$, i.e., the number of sensing nodes to keep after each sensing interval, is set in advance.  In our implementation and experiments, we set $k$ equals $0.5 * n$ as it produces the best results.

\section{Experiments}\label{sec:experiments}

In this section, we show the results of our experiments on real-world datasets to verify the effectiveness of our proposed approach.  We starts with describing our experimental settings (Section~\ref{sec:experiments}.A, and then, move on to compare our proposed approximation algorithm and bootstrapping strategies to the baselines (Section~\ref{sec:experiments}.B and \ref{sec:experiments}.C, respectively).  After that, we measure the effect of key parameters on the performance of our proposed algorithm (Section~\ref{sec:experiments}.D and \ref{sec:experiments}.E).

\subsection{Experimental settings}

In our experiments, we use a simulation-based approach using real-world datasets of opportunistic Bluetooth contacts.  Particularly, we use previously published datasets including (Table~\ref{fig:datasets}): i) mobile traces collected during SIGCOMM'09 \cite{pietilanen2012dissemination, crawdad} conference experiment in Barcelona, Spain, and ii) traces of Bluetooth encounters of a set of users collected through UIM experiment \cite{vu2010joint} at the University of Illinois at Urbana-Champaign in 2010.  For SIGCOMM'09 dataset, it was collected from 76 smartphones distributed to a set of volunteers during the first two days of the conference.  The participants were recruited on-site in conjunction of the conference registration.  Besides the logs of Bluetooth encounters, which represent all observed wireless contacts, each device was initialized with the social profile of the participant that included some basic information, such as list of friends and interests in the social profile.  For UIM dataset, it is collected by 28 Android phone users , who are staff, faculties, grads, and undergrads at University of Illinois, for 3 weeks in March 2010.

Since in the above datasets, all internal devices (i.e., 76 smartphones in SIGCOMM'09 and 28 Android phones in UIM) are doing the sensing task all the time (i.e., $\textbf{V}_{in}^{a} \equiv \textbf{V}_{in}$), we are able to have the ground-truth information about the best possible set of contacts we can collect (i.e., $\overline{\textbf{E}_{t_s}}$).  As a result, we are able to use the sensing coverage ratio (described in Section~\ref{sec:probdef}) as the \textit{evaluation metric}.  Basically, this ratio measures the ratio between the number of contacts that devices in $\textbf{V}_{in}^{a}$ can cover, and that by all devices in $\textbf{V}_{in}$.

In terms of \textit{algorithms}, we compare three approximation algorithms discussed in this paper: Top-n Random-based Approximation Vertex Cover (denoted as RANDOM), Top-n Greedy Vertex Cover (denoted as GREEDY), and our proposal -- Context-aware Approximation Algorithm (denoted as HCONTEXT).

For each experimental scenario, the time is divided into multiple rounds, each round consists of a sensing interval (i.e., $t_s$) and a decision interval (i.e., $t_d$) -- as describe in Section~\ref{sec:probdef}.  Since the main objective is to measure the sensing coverage performance of the approximation algorithms and bootstrapping methods, to simplify the experiment, we assume that all internal nodes have the same decision time interval $t_d$, and are be able to successfully send sensing data, receive sensing assignment from centralized server within $t_d$.  The experimental results are presented in graph with y-axis is the sensing coverage ratio and x-axis is the starting time of each sensing round.

\begin{figure}
\centering
\footnotesize
	\begin{tabular}{|c||c|c|}
			\hline   & SIGCOMM '09 & UIM \\ \hline
			\hline Context & Conference & University Campus \\
			\hline Duration (days) & 5 & 21 \\
			\hline Device type & Phone & Phone \\ 
			\hline $|\textbf{V}_{in}|$ & 76 & 28 \\ 
			\hline $|\textbf{V}_{ext}|$ & 11945 & 9015 \\
			\hline $\tau$ (seconds) & 120 & 60 \\ 
			\hline 
	\end{tabular}
\caption{Datasets}\label{fig:datasets}
\end{figure}

Since there are multiple parameters that could affect the performance of an algorithm (including sensing time interval $t_s$, number of active sensing nodes $n$, the selection of bootstrapping strategy $bootstrap(\textbf{V}_{in})$), in the following, we use a step-by-step experimenting approach.  Particularly, we start with comparing sensing coverage between different algorithms (Section~\ref{sec:experiments}.B) by fixing the configuration parameters (i.e., $t_s$, $n$, and $bootstrap(\textbf{V}_{in})$).  After that, we fix the algorithm, $t_s$, and $n$ to compare between bootstrapping strategies (Section~\ref{sec:experiments}.C).  Similarly, in Section~\ref{sec:experiments}.D and \ref{sec:experiments}.E, we measure the affect of $t_s$ and $n$ respectively by fixing the remaining parameters. 

\subsection{Coverage capability comparison}

\begin{figure*}[t!]
    \centering
    \begin{subfigure}[t]{0.3\textwidth}
        \centering
        \includegraphics[scale=0.45]{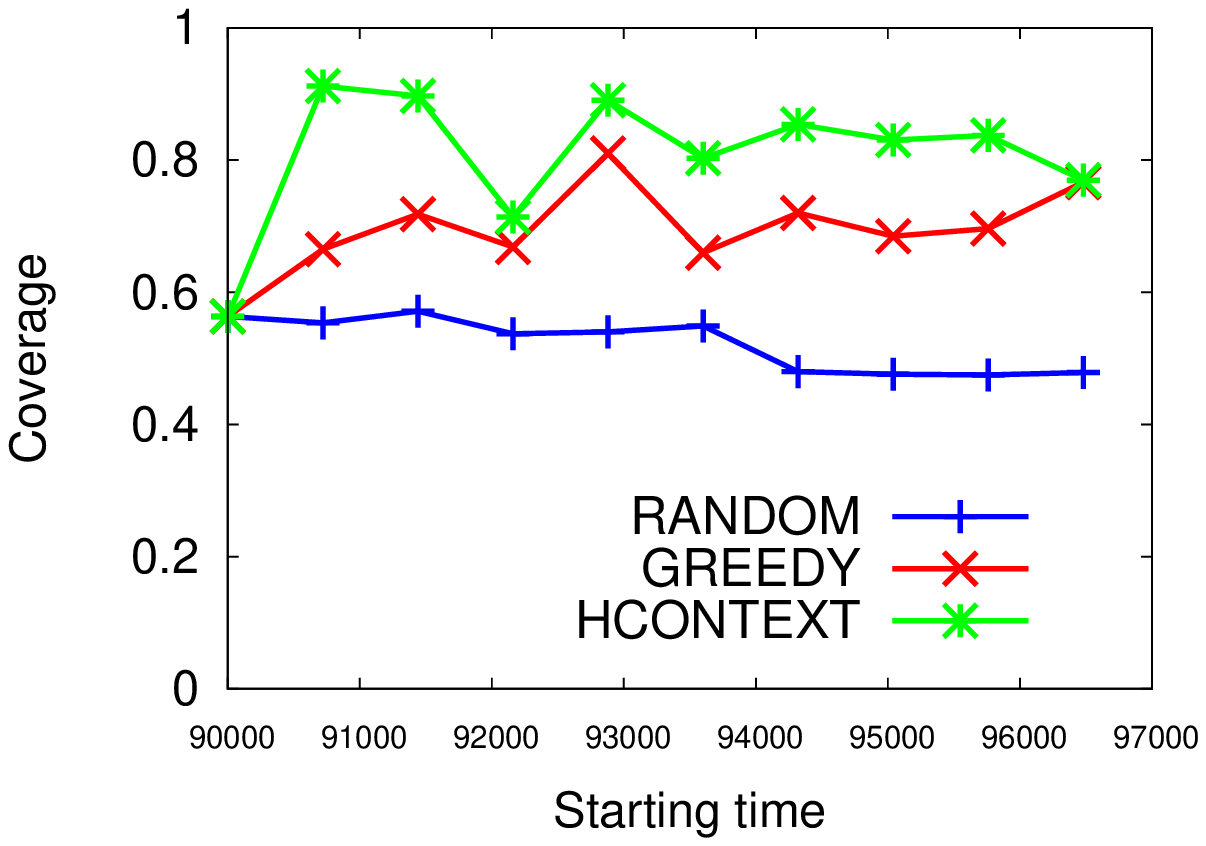}
        \caption{SIGCOMM'09 keynote session}
		\label{fig:coverage_1}
    \end{subfigure}\hfill%
    ~ 
    \begin{subfigure}[t]{0.3\textwidth}
        \centering
        \includegraphics[scale=0.45]{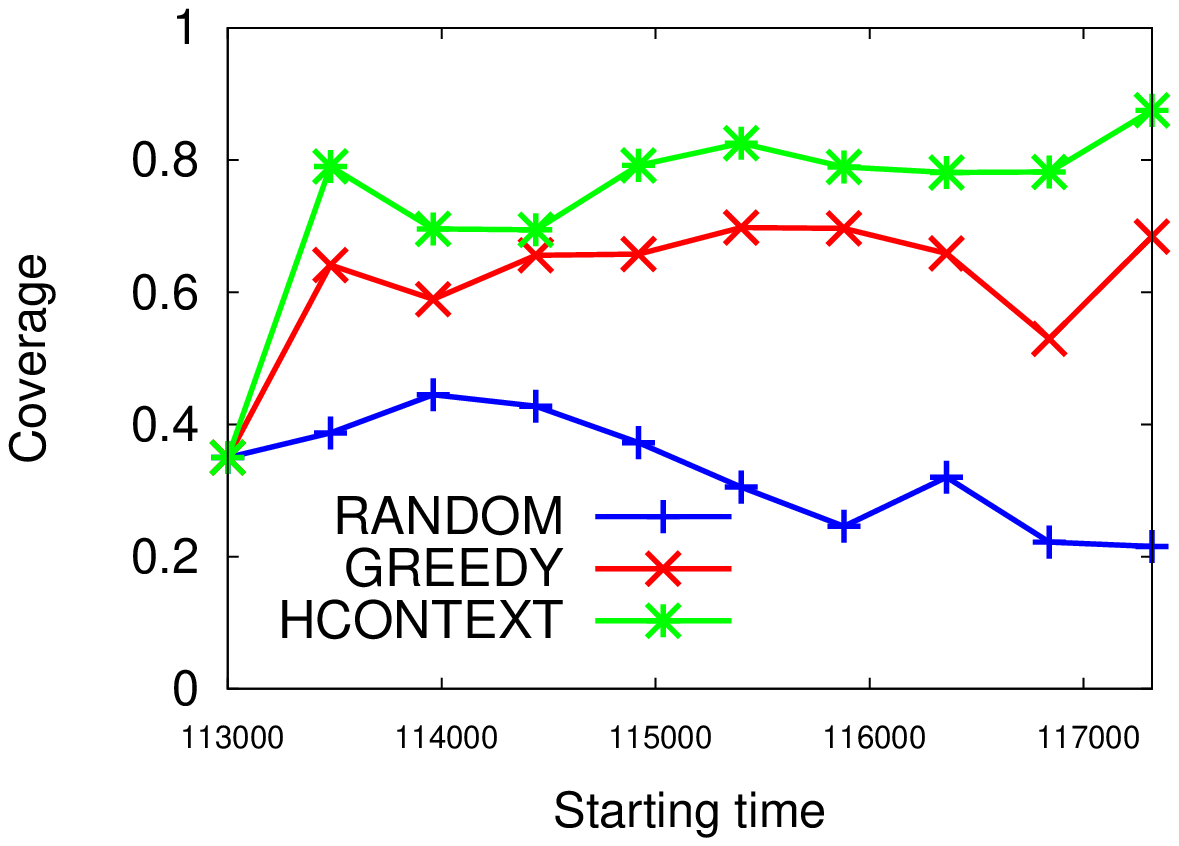}
		\caption{SIGCOMM'09 poster/demo session}
		\label{fig:coverage_2}
    \end{subfigure}\hfill%
    ~ 
    \begin{subfigure}[t]{0.3\textwidth}
        \centering
        \includegraphics[scale=0.45]{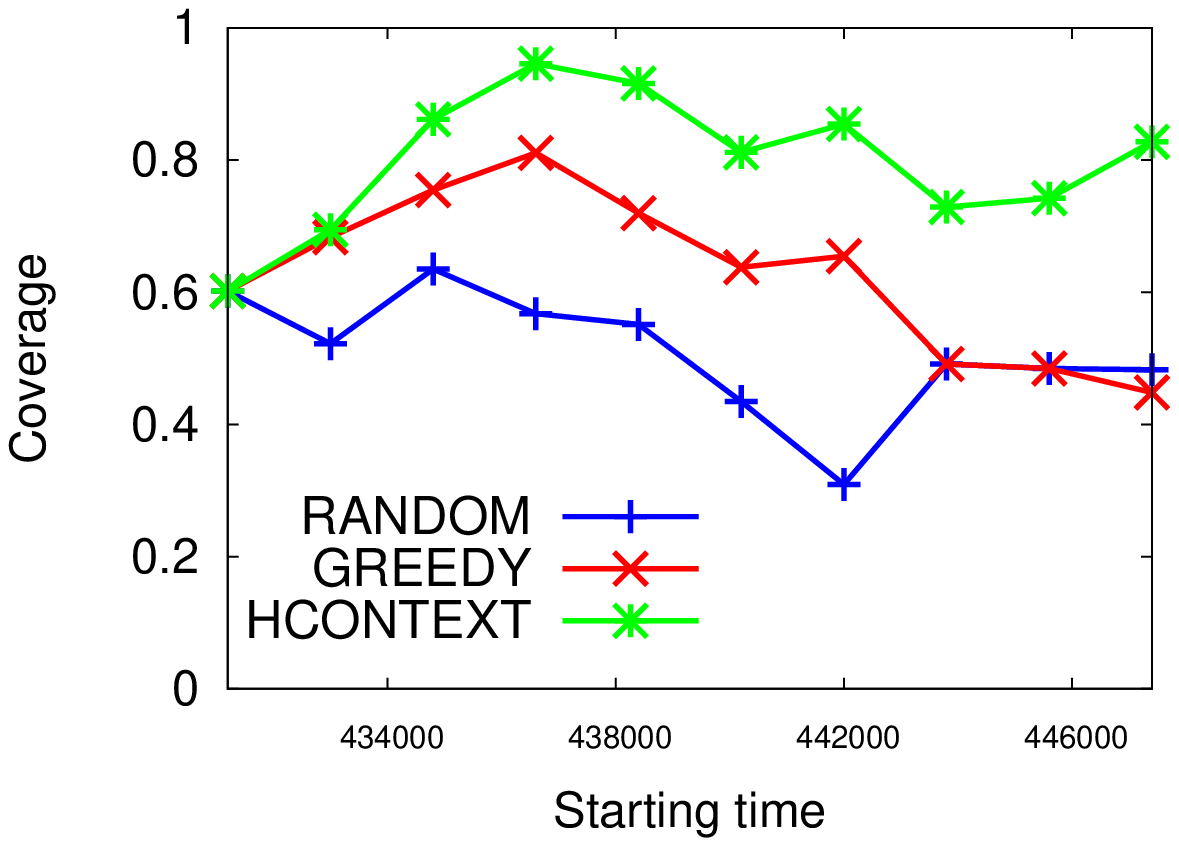}
		\caption{UIM's weekday interaction}
		\label{fig:coverage_3}
    \end{subfigure}
    \caption{Sensing coverage comparison}
\end{figure*}

In terms of coverage capability, we compare the three approximation algorithms on three different scenarios.  The first scenario, \textit{keynote presentation session} (Figure~\ref{fig:coverage_1}), is the duration when keynote presentations are undertaken and people gather into large conference room to listen to the presentations.  This scenario presents a crowd, less dynamic situation.  The second scenario, \textit{poster/demo and socialized session} (Figure~\ref{fig:coverage_2}), is the duration when the poster/demo session of the conference is undertaken. During this time, people gather in an open space venue and walk around to listen to poster/demo presentations.  This scenario represents a less crowd and more dynamic situation.  For the third scenario, we evaluate algorithms with the Bluetooth encounters of university users in UIM dataset during a weekday afternoon, where typical activities on campus are happening, such as classes, meeting, lab sessions, etc.

In terms of setting-up the paramters, for the length of sensing interval $t_s$, since the keynote presentation session is more crowd than the poster/demo session, we set the length of time interval shorter (i.e., 480s compared to 720s).  For the third scenario, since the daily encounters between university users happen at a lower frequency, the length of the sensing time interval is set to 30 minutes.  The effect of choosing different lengths of sensing time interval is studied in the next section.  For the sensing coverage comparison, we fix the bootstrapping strategy as random selection, the number of sensing devices is limited to be 40\% of total number of internal devices.

As we can see in Figure~\ref{fig:coverage_1}, \ref{fig:coverage_2}, and \ref{fig:coverage_3}, HCONTEXT outperforms RANDOM and GREEDY algorithm in all scenarios.  The RANDOM algorithm, although looks to be the simplest one, is still better than the more intuitive GREEDY algorithm.  However, it is not a surprise, since random-based strategy has been considered as one of the most effective general-purpose approximation  algorithms for vertex cover problem \cite{vertexcover, hochbaum1982approximation}.  The superior of HCONTEXT compared with two other algorithms is clearer in all scenarios, and this helps confirm our intuition of designing HCONTEXT that takes into account the contextual information in forms of node observability and coverage utility scores of sensing nodes.

\subsection{Bootstrapping methods comparison}

\begin{figure}[t!]
  	\centering
    \begin{subfigure}[t]{0.5\textwidth}
        \centering
        \includegraphics[scale=0.45]{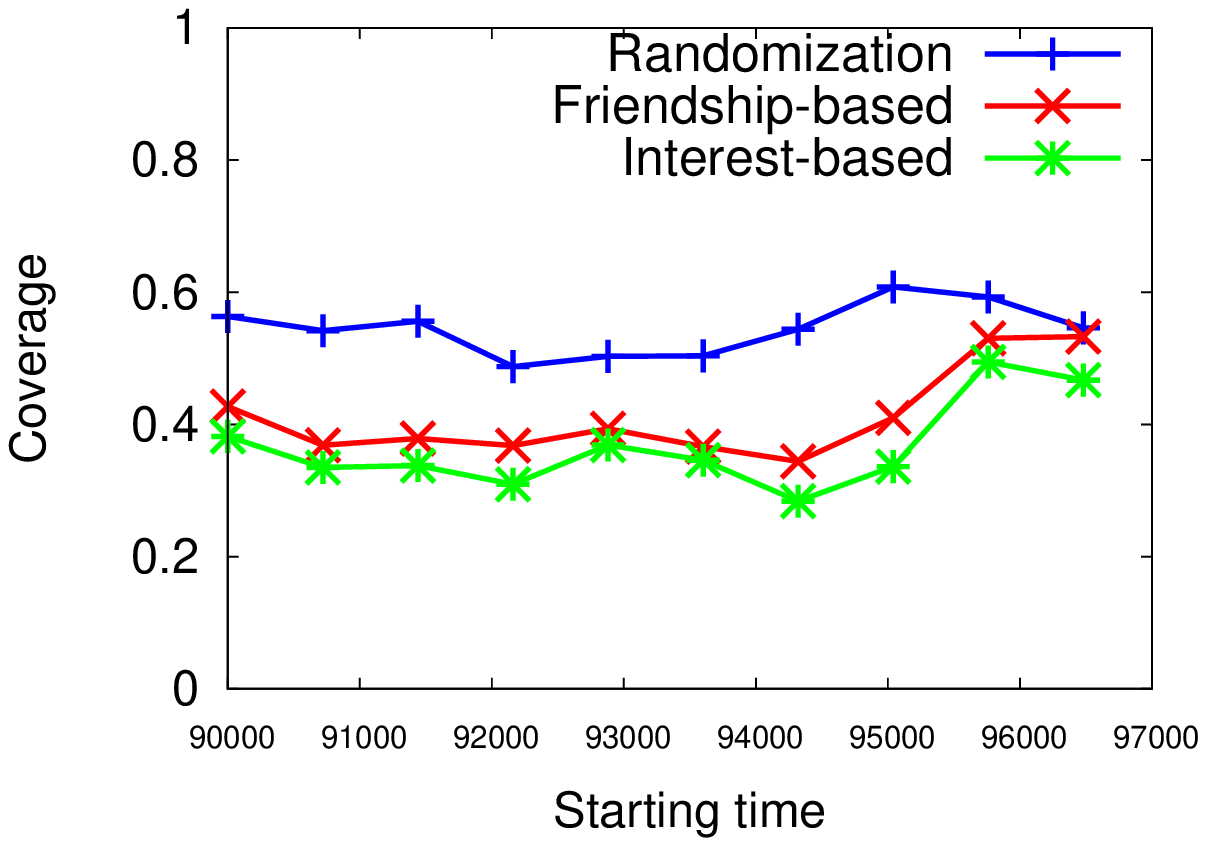}
        \caption{SIGCOMM'09 keynote session}
		\label{fig:bootstrap_1}
    \end{subfigure}
    \begin{subfigure}[t]{0.5\textwidth}
        \centering
        \includegraphics[scale=0.45]{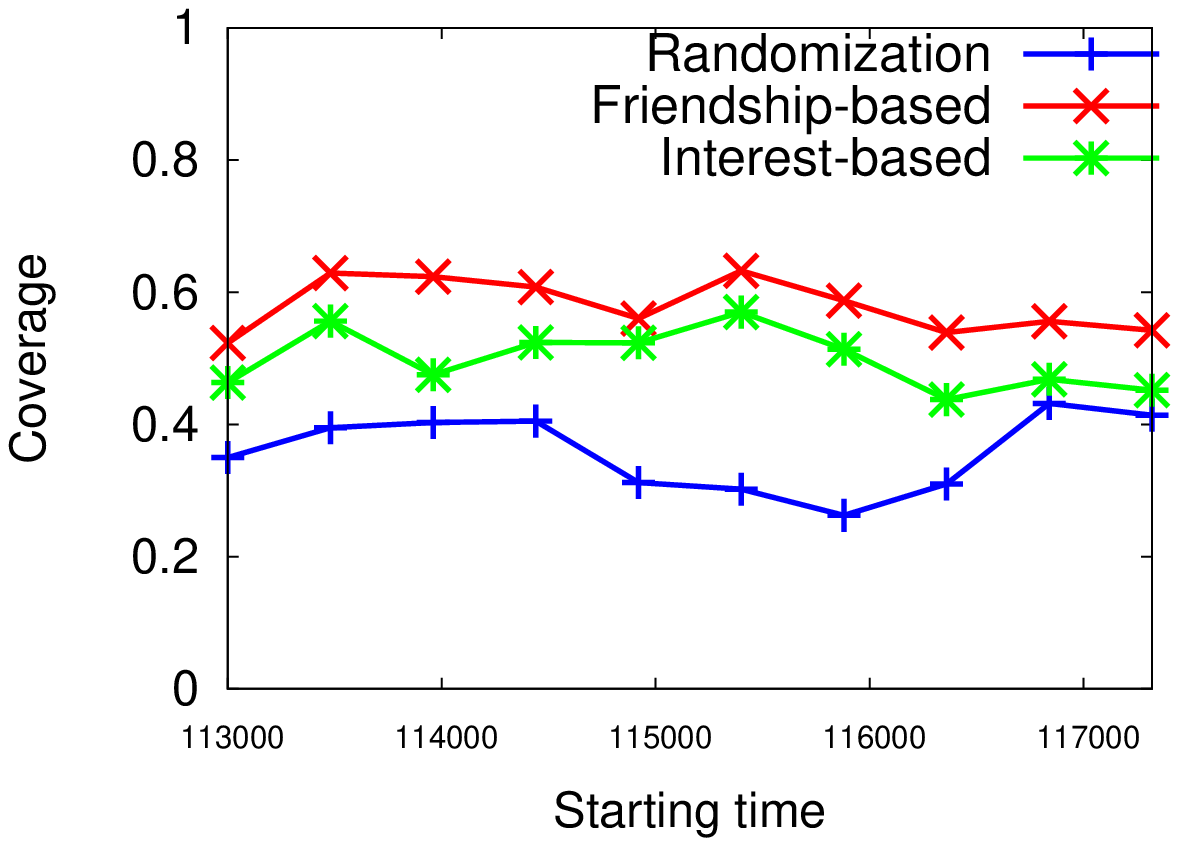}
		\caption{SIGCOMM'09 poster/demo session}
		\label{fig:bootstrap_2}
    \end{subfigure}
	\caption{Bootstrapping methods comparison}
\end{figure}

In this section, we compare three different bootstrapping methods for selecting initial set of sensing devices: i) Randomization, ii) Friendship-based, and iii) Interest-based bootstrapping.  We test each method with the HCONTEXT approximation algorithm during the SIGCOMM '09 keynote presentation session and poster/demo session, and report the sensing coverage for comparison (Figure~\ref{fig:bootstrap_1} and \ref{fig:bootstrap_2}).  Interestingly, the results show that during the keynote session, randomization-based bootstrapping yields the best initialized sensing coverage, followed by friendship-based, and lastly interest-based selection.  This is because, during such a crowd and less dynamic event, the wireless encounters tend to be more random (e.g., due to different arrival time of audience and seating arrangement), and are less influenced by the common interests or friendship.  On the other hand, during the poster and demo session, friendship-based bootstrapping produces the best result, followed by interest-based, and lastly randomization.  This result is reasonable, since during such a more socialized event, people tend to interact more with whom they share similar interests or they know in person (i.e., friends). 

In the next sections, we measure the effects of different parameters/constraints, including the length of sensing interval $t_s$ and the number of sensing devices $n$, to the performance of HCONTEXT algorithm.

\subsection{Varying different lengths of $t$}

\begin{figure}[!h]
\centering
\includegraphics[width=0.7\columnwidth]{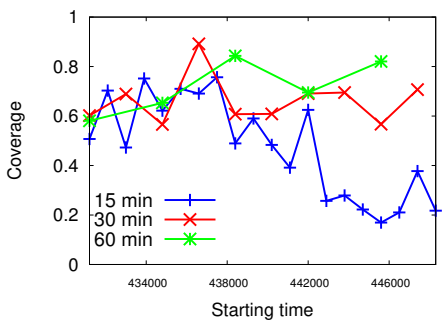}
\caption{Comparison of different lengths of time interval $t$}
\label{fig:intervals}
\end{figure}

Different length of a time interval affects the collection of wireless contacts by sensing devices.  Apparently, the longer the time interval, the more wireless contacts a node can observe.  If the sensing interval is set too short, the collected contact graph might not be complete, and thus, negatively affect our understanding of the context (i.e., node observability and coverage utility).

In our experiment, we measure the effect of the length of time interval to different scenarios.  The results support our aforementioned motivations.  For example, in university campus scenario, if the length of sensing time $t_s$ is set too short (i.e., 15 min), the result is not as good as when $t_s$ is set longer (i.e., 30 min or 60 min) (see Figure~\ref{fig:intervals}).  This is because 30 min or 60 min interval lengths are closer to the realistic frequency of people's encounters in daily life (while 15 min is too short).  This result suggests that, the length of sensing interval should be chosen carefully, based on the context of the environment that we want to sense. 

\subsection{Varying different number of sensing devices $n$}

\begin{figure}[!h]
\centering
\includegraphics[width=0.7\columnwidth]{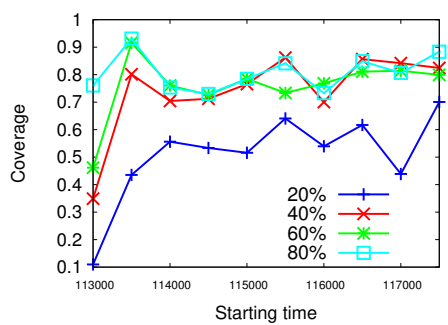}
\caption{Comparison of different percentages of sensing devices}
\label{fig:percentages}
\end{figure}

In our last experiment, we measure the effect of using different numbers of sensing devices $n$.  Obviously, the more sensing devices we have, the better coverage we could achieve.  However, as the number of sensing devices can be considered as a cost and energy constraints, we would like our method to still perform well when there are few sensing devices available.  In our experiment, we test our proposed HCONTEXT algorithm while varying the percentage of sensing devices in all internal devices during poster/demo session and report the coverage.  The result in Figure~\ref{fig:percentages} show that (not surprisingly) as the percentage increases, we are able to obtain better coverage.  More importantly, the result also shows an interesting insight about the ability of HCONTEXT to withstand the limited number of sensing devices.  At the lowest percentage (i.e., 20\%), although HCONTEXT starts not very well, it quickly gains better coverage and reaches a reasonably high coverage level, compared with higher percentage of sensing devices (see Figure~\ref{fig:percentages}).  With 40\% of sensing devices, the algorithm can perform almost as good as higher percentage levels.  This result is highly desirable as lower percentage of sensing devices means saving energy and cost.

In summary, throughout the experiments, we have shown that: \textit{i)} HCONTEXT is significantly better than other state-of-the-art approximation algorithms for crowd-sensing task;\textit{ ii)} Different bootstrapping strategies should be employed in different scenarios to obtain the best performance; \textit{iii)} The length of time interval needs to be set appropriately depending on the sensing context; and \textit{iv)} HCONTEXT still performs well with limited number of available sensing devices.
 
\section{Related Work}\label{sec:relatedwork}

The idea of  \textit{crowd computing} was first introduced by Murray et al. \cite{murray2010case}, which aims to combine mobile
devices and social interactions to achieve large-scale distributed computation.  The paper, however,  only proposed one realistic model for crowd computing: static task farming, which does not take into account the dynamic nature of the crowd.  In addition, while \cite{murray2010case} focus on computational resource, our optimization objective is in the coverage of the task.  \cite{kravets2013crowdwatch} presents CrowdWatch, a scalable, distributed and energy-efficient crowd-sourcing framework, based on a building a hierarchy of participants.  \cite{roitman2012harnessing} desribes a crowd sensing system that is developed in IBM for the smart cities domain that utilize heterogeneous types of data sources.  \cite{weppner2011collaborative} present a technique for estimating crowd density by using a mobile phone to scan the environment for discoverable Bluetooth devices.  Mashhadi et al. \cite{mashhadi2011quality} reasons on users mobility patterns and quality of their past contributions to estimate user's credibility under crowd-sensing scenario.  Along the line with human-centric computation, \cite{srivastava2012human} provides a comprehensive survey on human-centric sensing tasks.  Our paper falls nicely into the category of humans as sensor operators (collection campaigns) and directly solve one of the challenges mentioned in \cite{srivastava2012human}: to identify the appropriate set of individuals who would collect the data. Nicolai et al. present an interesting study \cite{nicolai2007relationship} that show limited number of people with discoverable Bluetooth devices (only 6\% devices in San Francisco are detectable).  This limitation does not affect our experiments, since the datasets used in our experiment, in fact, show that the number of external devices with discoverable Bluetooth during crowd event is quite big.

Sensor coverage problem \cite{meguerdichian2001exposure} has been long studied before, but usually in static setting (e.g., sensor placement).  Cardei et al. present a comprehensive survey \cite{cardei2006energy} on related work addressing energy-efficient coverage problems in the context of static wireless sensor networks.  In our paper, we are able to adapt with the mobility nature of  users by doing context-aware adjustment for assignment of sensing devices overtime.  Wu et al., \cite{wu1999calculating} model the problem of routing in wireless ad hoc network as the connected dominating set problem and propose distributed approximation algorithm to find the connected dominating set.  In our paper, we present a centralized sensing model for monitoring the crowd and do not require the sensing devices to be connected to each other.

Vertex cover problem \cite{vertexcover} is a well-know optimization problem in graph theory.  As far as we concern, this is the first paper that draw the connection between vertex cover and crowd-sensing problem.  Since finding minimum vertex cover is a NP-complete problem \cite{karp1972reducibility}, there have been efforts \cite{hochbaum1982approximation, dinur2005hardness} to develop approximation algorithms.  However, all of the proposed approximation algorithms are designed for generic graph.  In this paper, we propose a new approximation algorithm designed for crowd-sensing scenario that take into account the dynamic nature of the contact graph.

\section{Conclusion and future work}\label{sec:conclusions}

In summary, in this paper, we have modeled the crowd-sensing problem as an optimization problem and draw the connection to the vertex cover problem in graph theory.  We show that the current state-of-the-art approximation algorithms for vertex cover are not well-designed to deal with the dynamic nature of the crowd and its social and spatio-temporal characteristics.  We thus propose the notions of node observability and coverage utility score and design a new context-aware approximation algorithm and human-centric bootstrapping strategies to find vertex cover that is tailored for crowd-sensing task.  We have also verified the effectiveness of our proposed approach via comprehensive experiments on real-world datasets.

For future work, we would like to explore to include location information of the devices that can be inferred from Wifi access points data in the UIM dataset \cite{vu2010joint} to improve the assignment of sensing devices.  In addition, from the results on the varying the length of time intervals and percentage of sensing nodes, it is interesting to investigate new algorithms that can adaptively adjust the sensing interval length and decrease/increase the number of sensing nodes depending on the context of the crowd, while maintaining a desirable sensing coverage.

%
%
%
%
%
\balance


\bibliographystyle{hcontext}
\bibliography{hcontext.bib}
\end{document}